\documentclass[10pt,conference]{IEEEtran}

\IEEEoverridecommandlockouts
\usepackage{amsmath,amssymb,amsfonts}
\usepackage[backend=biber,style=ieee,natbib=true]{biblatex} 
\usepackage{algorithmic}
\usepackage{graphicx}
\usepackage{textcomp}
\usepackage{paralist}
\usepackage{booktabs}
\usepackage{mathtools}
\usepackage{multirow}
\usepackage[table]{xcolor}
\usepackage{makecell}
\usepackage{csquotes}
\usepackage{paralist}
\usepackage{siunitx}
\usepackage{graphicx}
\usepackage{amsmath}
\usepackage{listings}
\usepackage{mathtools}
\usepackage[para]{threeparttable}

\usepackage{tikz}

\def\BibTeX{{\rm B\kern-.05em{\sc i\kern-.025em b}\kern-.08em
    T\kern-.1667em\lower.7ex\hbox{E}\kern-.125emX}}

\makeatletter
\newenvironment{btHighlight}[1][]
{\begingroup\tikzset{bt@Highlight@par/.style={#1}}\begin{lrbox}{\@tempboxa}}
{\end{lrbox}\bt@HL@box[bt@Highlight@par]{\@tempboxa}\endgroup}

\newcommand\btHL[1][]{%
  \begin{btHighlight}[#1]\bgroup\aftergroup\bt@HL@endenv%
}
\def\bt@HL@endenv{%
  \end{btHighlight}%
  \egroup
}
\newcommand{\bt@HL@box}[2][]{%
  \tikz[#1]{%
    \pgfpathrectangle{\pgfpoint{1pt}{0pt}}{\pgfpoint{\wd #2}{\ht #2}}%
    \pgfusepath{use as bounding box}%
    \node[anchor=base west, fill=orange!30,outer sep=0pt,inner xsep=1pt, inner ysep=0pt, rounded corners=3pt, minimum height=\ht\strutbox+1pt,#1]{\raisebox{1pt}{\strut}\strut\usebox{#2}};
  }%
}
\makeatother
\def\benchsize{2,466}
\def\nexptypesinbench{37}

\def\throwbench{\textsc{ThrowBench}}
\def\downloadurl{\url{https://github.com/giganticode/throwbench/}}

\renewcommand{\bibfont}{\footnotesize} %
\begin{document}
\newcommand{\myparagraph}[1]{\noindent \textit{#1.}}

\lstdefinestyle{patch}{
    basicstyle=\footnotesize\ttfamily,
    moredelim=**[is][{\btHL[fill=green!30]}]{`}{`},
    moredelim=**[is][{\btHL[fill=red!30]}]{@}{@},
    escapeinside={(*}{*)},
    numbers=left,
    keywordstyle=\ttfamily\bfseries,
    stringstyle=\ttfamily\itshape
}

\newcommand*\circled[2]{\tikz[baseline=(char.base)]{
            \node[shape=circle,draw=#2,inner sep=1pt] (char) {#1};}}

\newcommand*\errorhere{\tikz[baseline=(char.base)]{
            \node[shape=circle,draw=black,fill=red,inner sep=1pt] (char) {!};}}

\title{\throwbench{}: Benchmarking LLMs by Predicting Runtime Exceptions}

\author{\IEEEauthorblockN{Julian Aron Prenner}
\IEEEauthorblockA{\textit{Free University of Bozen-Bolzano}\\
Bozen-Bolzano, Italy \\
prenner@inf.unibz.it}
\and
\IEEEauthorblockN{Romain Robbes}
\IEEEauthorblockA{\textit{Univ. Bordeaux, CNRS, Bordeaux INP},\\ \textit{LaBRI, UMR 5800} \\
Bordeaux, France \\
romain.robbes@u-bordeaux.fr}
}

\maketitle

\begin{abstract}
Modern Large Language Models (LLMs) have shown astounding capabilities of code understanding and synthesis. 
In order to assess such capabilities, several benchmarks have been devised (e.g., HumanEval). However, most benchmarks focus
on code synthesis from natural language instructions. Hence, such benchmarks do not test for other forms of code understanding. Moreover, there have been concerns about contamination and leakage. That is, benchmark problems (or closely related problems) may appear in training set, strongly biasing benchmark results. 
In this work we investigate whether large language models can correctly predict runtime program behavior. To this end, we introduce \throwbench{}, a benchmark consisting of over 2,400 short user-written programs written in four different programming languages. The majority of these programs throw an exception during runtime (due to a bug). LLMs are asked to predict whether a presented program throws an exception and, if so, which one. Evaluating our benchmark on six state-of-the-art code LLMs we see modest performance ranging from 19 to 38\,\% (F1 score).
Benchmarking a wider set of code capabilities could improve the assessment of code LLMs and help identify weak points in current models. Moreover, as ground-truth answers have been determined through program execution, leakage is not a concern.  We release \textsc{ThrowBench} as well as all of our results together with this work. 
\end{abstract}

\section{Introduction}
Large Language Models (LLMs) have established a new state of the art in several Software Engineering and code-related tasks, such as code summarization~\citep{ahmedFewshotTrainingLLMs2023}, code synthesis~\citep{austinProgramSynthesisLarge2021}, intelligent code completion~\citep{svyatkovskiyIntelliCodeComposeCode2020}, automatic code review~\citep{wadhwaCOREResolvingCode2024} or automated program repair~\citep{xiaKeepConversationGoing2023}.
In order to assess code capabilities of LLMs, they are subjected to various specifically designed benchmarks. Benchmark results allow for a direct comparison between different models and may expose an individual model's strengths and weaknesses.
One of the first such benchmarks was HumanEval introduced together with Codex~\citep{chenEvaluatingLargeLanguage2021}, one of the first bespoke code LLMs. 

Since then, many more benchmarks and benchmark extensions have been proposed, including MultiPL-E~\citep{cassanoMultiPLEScalablePolyglot2023}, MBPP~\citep{austinProgramSynthesisLarge2021}, LiveCodeBench~\citep{jainLiveCodeBenchHolisticContamination2024}, McEval~\citep{chaiMcEvalMassivelyMultilingual2024}, 
BigCodeBench~\citep{zhuoBigCodeBenchBenchmarkingCode2024}, EvalPlus~\citep{liuYourCodeGenerated2023} or CRUXEval~\citep{guCRUXEvalBenchmarkCode2024}.

In this work, we present yet another benchmark for evaluating LLMs named \throwbench{}. Unlike previous work, we focus on a novel benchmarking task: determining error runtime behavior of a program. That is, given a short program along with program input, we ask an LLM to predict whether the program will throw an exception (due to a bug in the code) and if so, which type of exception (e.g., \texttt{NullPointerException} or \texttt{StringIndexOutOfBoundsException} in Java). Doing so, requires a thorough understanding of code semantics and its behavior during runtime.

\paragraph*{Other benchmarks}
To the best of our knowledge predicting runtime exceptions is a novel benchmark task. Other aspects of our benchmark bear some similarity with previously proposed benchmarks, e.g., predicting runtime behavior (CRUXEval~\citep{guCRUXEvalBenchmarkCode2024}) or using submissions to programming contests (APPS~\citep{hendrycksMeasuringCodingChallenge2021}  or LiveCodeBench~\citep{jainLiveCodeBenchHolisticContamination2024}). We compare our benchmark to several other benchmarks and discuss differences and similarities in Section \ref{sec:background}.

\paragraph*{Approach}
The examples in our benchmark are short programs taken from the RunBugRun~\citep{prennerRunBugRunExecutableDataset2023} dataset (which in turn is based on CodeNet~\citep{puriCodeNetLargeScaleAI2021}). In particular, we select buggy programs that throw an exception during runtime (for a given test input). We evaluate \throwbench{} on six recent, state-of-the-art \enquote{open-weight} code LLMs. We detail these steps in Section~\ref{sec:approach}.

\paragraph*{Benchmark}
Our benchmark contains \benchsize{} example programs, written in Python, Java, C\# and Ruby throwing \nexptypesinbench{} different types of exceptions (for all languages combined).

\paragraph*{Evaluation}
Our evaluation shows performance differences across different LLMs, programming languages and exception types. Overall, performance is modest and ranges from 19 to 38\,\% (F1 score). We discuss evaluation results in Section~\ref{sec:results}.

Overall, \throwbench{} proves challenging for current state-of-the-art LLMs and there is ample space for improvements. The presented benchmark is intended to be used in \emph{addition} to other benchmarks and not as a replacement. The underlying idea is that testing different and multiple aspects of LLMs gives a more meaningful picture of a model's code capabilities. Our benchmark, as well as all model outputs are available at \downloadurl{}. 

\section{Background and Related Work}\label{sec:background}
In this section we will look at existing benchmarks for evaluating code LLMs and briefly discuss similarities and differences with respect to \throwbench{}.

\paragraph*{HumanEval}
A small dataset of 164 programming problems in Python introduced together with the Codex code LLM~\citep{chenEvaluatingLargeLanguage2021}. Each problem consists of a stub with a function signature and a docstring describing the behavior of the function but without actual implementation; test cases are provided to determine correctness. The model is asked to implement the corresponding function body.
HumanEval was extended in several ways by later work. For instance, \citet{cassanoMultiPLEScalablePolyglot2023} translated the problems in HumanEval into 18 additional programming languages (MultiPL-E). \citet{liuYourCodeGenerated2023} generated additional test cases for the problems in HumanEval to increase evaluation reliability.

\paragraph*{MBPP}
MBPP~\citep{austinProgramSynthesisLarge2021} is a dataset of 974 programming problems. A text description of the problem is provided together with test cases in the form of Python assert statements. The model is asked to implement a corresponding Python function.
Like HumanEval, MBPP was translated into additional programming languages (MultiPL-E)~\citep{cassanoMultiPLEScalablePolyglot2023}.

\paragraph*{CoNaLa}
CoNaLa~\citep{yinLearningMineAligned2018} is a dataset of parallel pairs of short natural language problems and code snippets in Python and Java. The problems have originally been mined from StackOverflow; a curated subset of 500 examples is commonly used as a benchmark.
In contrast with HumanEval or CoNaLa, the problems in CoNaLa are generally more \enquote{practical}; for instance, they may refer to specific usages of Python or Java APIs. \citet{wangMCoNaLaBenchmarkCode2023} translated the \emph{natural} language descriptions of CoNaLa into Spanish, Japanese and Russian (MCoNaLa).

\paragraph*{CodeXGLUE}
\citet{luCodeXGLUEMachineLearning2021} presented CodeXGLUE, a multi-task benchmark for code understanding. This benchmarks includes a wide range of different tasks, including code-to-code tasks such as clone detection, code translation or defect detection, text-to-code tasks such as code synthesis, code-to-text tasks such as code summarization and even a text-to-text task (documentation translation).

\paragraph*{LiveCodeBench}
LiveCodeBench~\citep{jainLiveCodeBenchHolisticContamination2024} is a benchmark based on problems taken from programming competition platforms such as AtCoder or LeetCode. 
LiveCodeBench is \enquote{live}, in the sense that as new problems are released on those platforms, they are also integrated into the benchmark. This means that the most recently added problems are guaranteed to be contamination-free if they have been released after the data cut-off date of a particular LLM.

\paragraph*{APPS}
APPS~\citep{hendrycksMeasuringCodingChallenge2021} is a large dataset of over 10,000 programming problems along with test cases and Python solutions taken from programming competitions (and thus, in this regard, bearing some similarity to LiveCodeBench~\citep{jainLiveCodeBenchHolisticContamination2024}).

\paragraph*{McEval}
McEval is a benchmark focusing on programming language diversity. It includes code in 40 different programming languages, including also rather \enquote{exotic} ones such as Tcl, AWK, VisualBasic, Lua or Julia. The benchmark contains code generation tasks (similar to e.g., MBPP), code explanation tasks, where code needs to be described in natural language and completion and fill-in tasks.

\paragraph*{BigCodeBench}
Another Python benchmark, similar in style to HumanEval or MBPP, with 1140 different problems. The benchmark comes in two types, a \enquote{code} flavor, where the problem description is given as a Python docstring (as in HumanEval) and an \enquote{instruct} flavor, where the problem description is given as a natural language prompt. 

\paragraph*{CRUXEval}
CRUXEval~\citep{guCRUXEvalBenchmarkCode2024} is a recent benchmark where the model must predict a Python function's input (i.e., function argument values) given its output (i.e., its return value) and vice versa. In other words, CRUXEval evaluates a model's capability to comprehend the runtime behavior of the function. The functions in CRUXEval are very short (3-13 lines) and have been generated by a LLM.

\paragraph*{ACES}
\citet{pourcelACESGeneratingDiverse2024} use LLMs to \emph{generate} a diverse set of programming puzzles (in Python) and find that the generated problems are harder to solve than existing benchmarks (e.g., HumanEval). Notably, problems are (also with the help of an LLM) labeled with skills required to solve them (e.g., \enquote{recursion} or \enquote{dynamic programming}), allowing a per-label evaluation of LLMs.

\paragraph*{This work}
The benchmark proposed in this work contains aspects already present in some of the above-mentioned benchmarks but to the best of our knowledge its central idea (predicting exceptions) is novel. First, like CRUXEval~\citep{guCRUXEvalBenchmarkCode2024} our benchmark tests for runtime behavior. However, while in CRUXEval program input and output must be predicted, in \throwbench{} the model must predict what type of exception is thrown. Unlike CRUXEval, \throwbench{} is, similar to McEval or MultiPL-E, a multilingual benchmark, covering four different programming languages. Moreover, examples in \throwbench{} have been written by humans and generally more complex and longer than those in CRUXEval (where examples have been generated by a LLM).
Next, like LiveCodeBench~\citep{jainLiveCodeBenchHolisticContamination2024}, \throwbench{} was designed with concerns about contamination in mind. As the code in training corpora hardly contains any information on their runtime behavior, we believe that \throwbench{} is (currently) free from any contamination issues.
Finally, like LiveCodeBench~\citep{jainLiveCodeBenchHolisticContamination2024} and APPS~\citep{hendrycksMeasuringCodingChallenge2021}, \throwbench{} is based on submissions to programming contests. More concretely, the examples in this benchmark have been taken from the RunBugRun~\citep{prennerRunBugRunExecutableDataset2023} dataset which in turn is based on IBM's CodeNet dataset~\citep{puriCodeNetLargeScaleAI2021}, a very large dataset of over 14 million submissions to several programming contests written in over 55 programming languages. RunBugRun is a filtered subset of CodeNet; it contains parallel pairs of buggy and fixed submissions. In particular, all programs in RunBugRun are syntactically correct, compilable and executable. This is what allowed us to select (buggy) programs throwing exceptions during runtime.

\begin{figure*}
    \centering
    \fontfamily{cmtt}\fontsize{7}{8}\selectfont
\begin{lstlisting}[breaklines=true,breakautoindent=false,breakindent=0ex]
You are given a Python program that either does not raise an exception or raises one of the following exceptions:
   NameError, TypeError, IndexError, ValueError, AttributeError, ZeroDivisionError,
   KeyError, ModuleNotFoundError, UnboundLocalError
Please find and output the name of the exception (or output "no_exception" if no exception is raised).
Give your answer in the following format: "Answer: <exception_name>" or "Answer: no_exception" (if no exception is raised).
Only output your answer. Do not output explanations.

Code:
----------------
<CODE>
----------------

Program Input:
----------------
<INPUT>
----------------
\end{lstlisting}
\caption{Prompt used in our evaluation. Following common conventions, we use the verb \emph{raise} for Python and Ruby, and \emph{throw} for C\# and Java. \texttt{<CODE>} is replaced with the example code, \texttt{<INPUT>} with the input that triggers the exception.}
\label{fig:prompt}
\end{figure*}

\begin{figure}
  \begin{lstlisting}[language=Python,style=patch,numbers=none,xleftmargin=0pt,xrightmargin=0pt,framesep=0pt, breaklines=true,postbreak=\mbox{\textcolor{red}{$\hookrightarrow$}\space},]
A=list(input())
n=len(A)
if(A[0]!="A"):print("WA")
elif(A.count("C")!=1):print("WA")
elif(A.index("C")<2 or A.index("C")>n-2):print("WA")
else:
  A.pop(A.index("C"))
  A.pop(0)
  if(A.islower()):print("AC") (*\errorhere{}*)
  else:print("WA")
\end{lstlisting}
\caption{A simple Python example in \throwbench{}. Independent of input, variable \texttt{A} is of type list and does not have a method \texttt{islower}, resulting in a \texttt{AttributeError} being thrown.}
\label{fig:example-py}
\end{figure}

\begin{figure}
  \begin{lstlisting}[language=Java,style=patch,numbers=none,xleftmargin=0pt,xrightmargin=0pt,framesep=0pt, breaklines=true,postbreak=\mbox{\textcolor{red}{$\hookrightarrow$}\space},]
Scanner sc = new Scanner(System.in);
long n = sc.nextLong();
long m = sc.nextLong();
String ans = "No";
if ((n / m) == 1) { (*\errorhere{}*)
  ans = "Yes";
}
System.out.println(ans);
\end{lstlisting}
\caption{A simple Java example in \throwbench{}. On input \texttt{-1 0}, an \texttt{ArithmeticException} is thrown. Note: imports, main class and main method omitted for brevity.}
\label{fig:example-java}
\end{figure}

\section{Approach}\label{sec:approach}

This section will outline, first, how \throwbench{} was created by selecting bugs from the RunBugRun~\citep{prennerRunBugRunExecutableDataset2023} dataset and, second, how we evaluated our benchmark on six state-of-the-art code LLMs.

\subsection{Creation}
The creation of \throwbench{} involved the following steps:

\begin{itemize}
    \item We select buggy programs in RunBugRun~\citep{prennerRunBugRunExecutableDataset2023} written in either Python, Ruby, C\# or Java as these are all languages that use exceptions.
    \item We execute the selected programs on the corresponding test cases (also part of RunBugRun), relying on RunBugRun's execution sandbox.
    \item We only keep programs that throw an exception during runtime.
    \item We subsample to balance the data in terms of both, programming languages and exception types.
    \item We discard very long programs; more concretely, we discard programs longer than 53 lines of code (95\textsuperscript{th} percentile).
    \item For each language, we add 10\,\% of correct programs (that is, programs that do not throw any exception on test input). The correct programs are, again, taken from RunBugRun. 
\end{itemize}

After these steps, we obtain a dataset of \benchsize{} programs (471 in C\#, 591 in Java,  834 in Python and 570 in Ruby), raising \nexptypesinbench{} different types of exceptions (see Table~\ref{tab:results}). Figures~\ref{fig:example-py} and \ref{fig:example-java} show two examples from the benchmark. The median program length is 18 lines of code.

\subsection{Evaluation}
To evaluate \throwbench{}, we rely on Ollama~\citep{ollama}, a \textsl{llama.cpp}-based~\citep{gerganovGgerganovLlamacpp2024} program facilitating the download and use of LLMs. 
We use the prompt shown in Figure~\ref{fig:prompt}; importantly, for each language, we list the possible exceptions as options (this makes the task slightly easier). Relying on Ollama's Python API, all examples for the benchmark are evaluated on each of the six \enquote{open-weight} code LLMs listed in Table~\ref{tab:models} whose sizes range from 7 to 34 billion parameters. In particular, we consider the last occurrence of an exception type (or the symbol \texttt{no\_exception}) in the model output as the predicted exception type. Evaluation results are presented and discussed in Section~\ref{sec:results}.

\begin{table}[h]
    \centering
    \setlength{\tabcolsep}{3pt}
    \caption{Evaluation results. Precision, recall and F1 is given as micro average.}
    \label{tab:models}
    \begin{tabular}{rrcccc}
        \thead{\textbf{Model}} & \thead{\textbf{Size}} & \thead{\textbf{Quant.}} & \thead{\textbf{Prec.}} & \thead{\textbf{Rec.}} & \thead{\textbf{F1}} \\
        \toprule
        \makecell[r]{CodeGemma Instr.~\citep{teamCodeGemmaOpenCode2024}} & 7B & FP16 &
        21.8 & 20.5 & 21.1 \\
        \makecell[r]{DeepSeek Coder 2 Instr.~\citep{deepseek-aiDeepSeekCoderV2BreakingBarrier2024}} & 16B & Q4\_0 &
        20.4 & 20.2 & 20.3 \\
        \makecell[r]{Codestral~\citep{aiCodestralHelloWorld2024}} & 22B & Q4\_0 &
        31.0 & 30.3 & 30.6 \\
        \makecell[r]{Qwen2.5 Coder Instr.~\citep{huiQwen25CoderTechnicalReport2024}} & 32B & Q4\_K\_M &
        \textbf{40.8} & \textbf{35.9} & \textbf{38.2} \\        
        \makecell[r]{DeepSeek Coder Instr.~\citep{guoDeepSeekCoderWhenLarge2024}} & 33B & Q4\_0 &
        27.9 & 26.5 & 27.2 \\        
        \makecell[r]{CodeLlama Instr.~\citep{roziereCodeLlamaOpen2024}} & 34B & Q4\_0 &
        19.9 & 18.2 & 19.0 \\
        \bottomrule
    \end{tabular}  
\end{table}

\begin{table}[h]
\begin{threeparttable}
    \centering
    \footnotesize
    \setlength{\tabcolsep}{1pt}
    \caption{F1 scores per language and exception type.}
    \label{tab:results}
    \begin{tabular}{rrrrrrrrr}
        & \thead{\textbf{Exception}\tnote{1}} & \thead{\textbf{CG}\tnote{2}} & \thead{\textbf{DC2}\tnote{3}} & \thead{\textbf{CS}\tnote{4}} & \thead{\textbf{QC}\tnote{5}} & \thead{\textbf{DC}\tnote{6}} & \thead{\textbf{CL}\tnote{7}} & \thead{\textbf{\#}} \\
        \toprule

\multirow{10}{*}{C\#} & ArgumentNull & \cellcolor[rgb]{0.969,0.988,0.961} 0.0 & \cellcolor[rgb]{0.969,0.988,0.961} 0.0 & \cellcolor[rgb]{0.969,0.988,0.961} 0.0 & \cellcolor[rgb]{0.969,0.988,0.961} 0.0 & \cellcolor[rgb]{0.969,0.988,0.961} 0.0 & \cellcolor[rgb]{0.969,0.988,0.961} 0.0 & 9\\
 & ArgumentOutOfRange & \cellcolor[rgb]{0.667,0.866,0.643} 27.5 & \cellcolor[rgb]{0.951,0.981,0.94} 2.6 & \cellcolor[rgb]{0.507,0.793,0.506} 37.1 & \cellcolor[rgb]{0.463,0.773,0.47} \textbf{39.7} & \cellcolor[rgb]{0.574,0.824,0.561} 33.3 & \cellcolor[rgb]{0.911,0.966,0.894} 8.4 & 76\\
 & KeyNotFound & \cellcolor[rgb]{0.794,0.919,0.768} 19.0 & \cellcolor[rgb]{0.969,0.988,0.961} 0.0 & \cellcolor[rgb]{0.672,0.868,0.647} 27.3 & \cellcolor[rgb]{0.546,0.811,0.538} \textbf{34.8} & \cellcolor[rgb]{0.894,0.959,0.874} 10.5 & \cellcolor[rgb]{0.969,0.988,0.961} 0.0 & 18\\
 & DivideByZero & \cellcolor[rgb]{0.728,0.892,0.702} 23.6 & \cellcolor[rgb]{0.89,0.958,0.87} 10.9 & \cellcolor[rgb]{0.117,0.528,0.256} \textcolor{white}{61.4} & \cellcolor[rgb]{0.0,0.378,0.152} \textcolor{white}{\textbf{72.9}} & \cellcolor[rgb]{0.179,0.589,0.304} \textcolor{white}{56.5} & \cellcolor[rgb]{0.333,0.709,0.403} 45.9 & 51\\
 & Format & \cellcolor[rgb]{0.831,0.934,0.807} 15.6 & \cellcolor[rgb]{0.922,0.97,0.907} 6.8 & \cellcolor[rgb]{0.686,0.874,0.661} 26.5 & \cellcolor[rgb]{0.861,0.946,0.839} 13.3 & \cellcolor[rgb]{0.618,0.845,0.597} \textbf{30.8} & \cellcolor[rgb]{0.824,0.931,0.799} 16.3 & 56\\
 & IndexOutOfRange & \cellcolor[rgb]{0.602,0.837,0.583} 31.7 & \cellcolor[rgb]{0.911,0.966,0.894} 8.2 & \cellcolor[rgb]{0.48,0.78,0.483} 38.5 & \cellcolor[rgb]{0.289,0.687,0.381} \textbf{48.2} & \cellcolor[rgb]{0.507,0.793,0.506} 36.9 & \cellcolor[rgb]{0.924,0.971,0.909} 6.5 & 88\\
 & InvalidOperation & \cellcolor[rgb]{0.879,0.953,0.858} 11.8 & \cellcolor[rgb]{0.969,0.988,0.961} 0.0 & \cellcolor[rgb]{0.718,0.888,0.693} \textbf{24.2} & \cellcolor[rgb]{0.864,0.947,0.843} 12.9 & \cellcolor[rgb]{0.969,0.988,0.961} 0.0 & \cellcolor[rgb]{0.92,0.969,0.904} 6.9 & 28\\
 & NullReference & \cellcolor[rgb]{0.969,0.988,0.961} 0.0 & \cellcolor[rgb]{0.969,0.988,0.961} 0.0 & \cellcolor[rgb]{0.92,0.969,0.904} \textbf{7.1} & \cellcolor[rgb]{0.969,0.988,0.961} 0.0 & \cellcolor[rgb]{0.92,0.969,0.904} \textbf{7.1} & \cellcolor[rgb]{0.969,0.988,0.961} 0.0 & 26\\
 & Overflow & \cellcolor[rgb]{0.469,0.775,0.474} 39.3 & \cellcolor[rgb]{0.969,0.988,0.961} 0.0 & \cellcolor[rgb]{0.402,0.742,0.437} 42.7 & \cellcolor[rgb]{0.463,0.773,0.47} 39.6 & \cellcolor[rgb]{0.376,0.73,0.424} \textbf{44.0} & \cellcolor[rgb]{0.775,0.911,0.747} 20.5 & 77\\
 & \textsl{No exception} & \cellcolor[rgb]{0.907,0.964,0.888} 9.0 & \cellcolor[rgb]{0.824,0.931,0.799} 16.5 & \cellcolor[rgb]{0.761,0.905,0.734} 21.4 & \cellcolor[rgb]{0.718,0.888,0.693} \textbf{24.2} & \cellcolor[rgb]{0.756,0.903,0.729} 21.8 & \cellcolor[rgb]{0.805,0.924,0.78} 18.1 & 42\\
 & \textsl{Avg (micro)} & \cellcolor[rgb]{0.737,0.896,0.711} 22.9 & \cellcolor[rgb]{0.89,0.958,0.87} 10.7 & \cellcolor[rgb]{0.602,0.837,0.583} 31.6 & \cellcolor[rgb]{0.552,0.814,0.542} \textbf{34.6} & \cellcolor[rgb]{0.624,0.847,0.602} 30.4 & \cellcolor[rgb]{0.82,0.93,0.795} 16.7 & -\\
\midrule
\multirow{11}{*}{Java} & Arithmetic & \cellcolor[rgb]{0.53,0.804,0.524} 35.6 & \cellcolor[rgb]{0.175,0.585,0.301} \textcolor{white}{56.6} & \cellcolor[rgb]{0.078,0.494,0.228} \textcolor{white}{64.2} & \cellcolor[rgb]{0.42,0.752,0.446} 41.8 & \cellcolor[rgb]{0.001,0.428,0.173} \textcolor{white}{\textbf{69.8}} & \cellcolor[rgb]{0.227,0.641,0.342} \textcolor{white}{52.5} & 71\\
 & ArrayIndexOutOfBounds & \cellcolor[rgb]{0.563,0.819,0.552} 33.9 & \cellcolor[rgb]{0.618,0.845,0.597} 30.8 & \cellcolor[rgb]{0.491,0.785,0.492} \textbf{38.0} & \cellcolor[rgb]{0.77,0.909,0.743} 20.8 & \cellcolor[rgb]{0.552,0.814,0.542} 34.4 & \cellcolor[rgb]{0.644,0.856,0.62} 29.3 & 80\\
 & IndexOutOfBounds & \cellcolor[rgb]{0.969,0.988,0.961} 0.0 & \cellcolor[rgb]{0.351,0.718,0.412} \textbf{45.0} & \cellcolor[rgb]{0.474,0.778,0.479} 38.9 & \cellcolor[rgb]{0.969,0.988,0.961} 0.0 & \cellcolor[rgb]{0.746,0.899,0.72} 22.2 & \cellcolor[rgb]{0.864,0.947,0.843} 12.9 & 29\\
 & NegativeArraySize & \cellcolor[rgb]{0.969,0.988,0.961} 0.0 & \cellcolor[rgb]{0.345,0.715,0.409} 45.5 & \cellcolor[rgb]{0.138,0.546,0.271} \textcolor{white}{\textbf{60.0}} & \cellcolor[rgb]{0.82,0.93,0.795} 16.7 & \cellcolor[rgb]{0.42,0.752,0.446} 41.7 & \cellcolor[rgb]{0.681,0.872,0.656} 26.7 & 11\\
 & NullPointer & \cellcolor[rgb]{0.931,0.974,0.917} 5.4 & \cellcolor[rgb]{0.931,0.974,0.917} 5.4 & \cellcolor[rgb]{0.931,0.974,0.917} 5.4 & \cellcolor[rgb]{0.931,0.974,0.917} 5.4 & \cellcolor[rgb]{0.816,0.928,0.791} \textbf{17.0} & \cellcolor[rgb]{0.969,0.988,0.961} 0.0 & 36\\
 & NumberFormat & \cellcolor[rgb]{0.913,0.967,0.896} 8.1 & \cellcolor[rgb]{0.658,0.862,0.633} 28.3 & \cellcolor[rgb]{0.519,0.798,0.515} \textbf{36.4} & \cellcolor[rgb]{0.935,0.975,0.922} 4.8 & \cellcolor[rgb]{0.552,0.814,0.542} 34.4 & \cellcolor[rgb]{0.742,0.897,0.715} 22.7 & 66\\
 & StackOverflow & \cellcolor[rgb]{0.969,0.988,0.961} 0.0 & \cellcolor[rgb]{0.969,0.988,0.961} 0.0 & \cellcolor[rgb]{0.969,0.988,0.961} 0.0 & \cellcolor[rgb]{0.969,0.988,0.961} 0.0 & \cellcolor[rgb]{0.969,0.988,0.961} 0.0 & \cellcolor[rgb]{0.969,0.988,0.961} 0.0 & 5\\
 & StringIndexOutOfBounds & \cellcolor[rgb]{0.289,0.687,0.381} 48.3 & \cellcolor[rgb]{0.469,0.775,0.474} 39.3 & \cellcolor[rgb]{0.245,0.66,0.357} \textcolor{white}{50.9} & \cellcolor[rgb]{0.918,0.968,0.901} 7.2 & \cellcolor[rgb]{0.23,0.645,0.345} \textcolor{white}{\textbf{52.1}} & \cellcolor[rgb]{0.53,0.804,0.524} 35.8 & 80\\
 & InputMismatch & \cellcolor[rgb]{0.929,0.973,0.914} 5.8 & \cellcolor[rgb]{0.969,0.988,0.961} 0.0 & \cellcolor[rgb]{0.794,0.919,0.768} 18.9 & \cellcolor[rgb]{0.253,0.668,0.363} \textbf{50.0} & \cellcolor[rgb]{0.502,0.791,0.501} 37.3 & \cellcolor[rgb]{0.861,0.946,0.839} 13.2 & 80\\
 & NoSuchElement & \cellcolor[rgb]{0.846,0.94,0.823} 14.6 & \cellcolor[rgb]{0.875,0.952,0.854} 12.1 & \cellcolor[rgb]{0.872,0.95,0.85} 12.4 & \cellcolor[rgb]{0.723,0.89,0.697} \textbf{24.0} & \cellcolor[rgb]{0.924,0.971,0.909} 6.3 & \cellcolor[rgb]{0.935,0.975,0.922} 4.8 & 80\\
 & \textsl{No exception} & \cellcolor[rgb]{0.846,0.94,0.823} 14.5 & \cellcolor[rgb]{0.787,0.916,0.76} 19.4 & \cellcolor[rgb]{0.77,0.909,0.743} 20.9 & \cellcolor[rgb]{0.634,0.852,0.611} \textbf{30.0} & \cellcolor[rgb]{0.709,0.884,0.684} 24.8 & \cellcolor[rgb]{0.779,0.913,0.752} 20.2 & 53\\
 & \textsl{Avg (micro)} & \cellcolor[rgb]{0.765,0.907,0.738} 21.1 & \cellcolor[rgb]{0.704,0.882,0.679} 25.3 & \cellcolor[rgb]{0.618,0.845,0.597} 30.8 & \cellcolor[rgb]{0.69,0.876,0.665} 26.1 & \cellcolor[rgb]{0.541,0.809,0.533} \textbf{35.3} & \cellcolor[rgb]{0.723,0.89,0.697} 23.8 & -\\
\midrule
\multirow{12}{*}{Python} & Attribute & \cellcolor[rgb]{0.913,0.967,0.896} 8.0 & \cellcolor[rgb]{0.639,0.854,0.615} 29.5 & \cellcolor[rgb]{0.264,0.675,0.369} 49.5 & \cellcolor[rgb]{0.0,0.418,0.169} \textcolor{white}{\textbf{70.6}} & \cellcolor[rgb]{0.915,0.968,0.899} 7.7 & \cellcolor[rgb]{0.913,0.967,0.896} 8.0 & 72\\
 & Index & \cellcolor[rgb]{0.681,0.872,0.656} 26.7 & \cellcolor[rgb]{0.779,0.913,0.752} 20.2 & \cellcolor[rgb]{0.602,0.837,0.583} 31.8 & \cellcolor[rgb]{0.168,0.578,0.295} \textcolor{white}{\textbf{57.4}} & \cellcolor[rgb]{0.662,0.864,0.638} 27.9 & \cellcolor[rgb]{0.794,0.919,0.768} 18.8 & 72\\
 & Key & \cellcolor[rgb]{0.791,0.918,0.764} 19.3 & \cellcolor[rgb]{0.507,0.793,0.506} 37.1 & \cellcolor[rgb]{0.574,0.824,0.561} 33.3 & \cellcolor[rgb]{0.345,0.715,0.409} \textbf{45.4} & \cellcolor[rgb]{0.77,0.909,0.743} 20.9 & \cellcolor[rgb]{0.875,0.952,0.854} 12.0 & 72\\
 & Name & \cellcolor[rgb]{0.737,0.896,0.711} 22.9 & \cellcolor[rgb]{0.53,0.804,0.524} 35.9 & \cellcolor[rgb]{0.249,0.664,0.36} \textcolor{white}{50.6} & \cellcolor[rgb]{0.013,0.439,0.182} \textcolor{white}{\textbf{69.0}} & \cellcolor[rgb]{0.502,0.791,0.501} 37.2 & \cellcolor[rgb]{0.861,0.946,0.839} 13.3 & 72\\
 & Overflow & \cellcolor[rgb]{0.875,0.952,0.854} 12.1 & \cellcolor[rgb]{0.969,0.988,0.961} 0.0 & \cellcolor[rgb]{0.872,0.95,0.85} 12.2 & \cellcolor[rgb]{0.695,0.878,0.67} \textbf{25.9} & \cellcolor[rgb]{0.969,0.988,0.961} 0.0 & \cellcolor[rgb]{0.918,0.968,0.901} 7.3 & 39\\
 & Recursion & \cellcolor[rgb]{0.32,0.702,0.397} \textbf{46.9} & \cellcolor[rgb]{0.783,0.915,0.756} 19.8 & \cellcolor[rgb]{0.507,0.793,0.506} 37.0 & \cellcolor[rgb]{0.427,0.755,0.449} 41.3 & \cellcolor[rgb]{0.681,0.872,0.656} 26.7 & \cellcolor[rgb]{0.629,0.85,0.606} 30.1 & 72\\
 & Syntax & \cellcolor[rgb]{0.704,0.882,0.679} 25.0 & \cellcolor[rgb]{0.732,0.894,0.706} 23.3 & \cellcolor[rgb]{0.491,0.785,0.492} 38.0 & \cellcolor[rgb]{0.0,0.388,0.156} \textcolor{white}{\textbf{72.4}} & \cellcolor[rgb]{0.704,0.882,0.679} 25.0 & \cellcolor[rgb]{0.913,0.967,0.896} 8.0 & 72\\
 & Type & \cellcolor[rgb]{0.85,0.941,0.827} 14.3 & \cellcolor[rgb]{0.813,0.927,0.787} 17.4 & \cellcolor[rgb]{0.474,0.778,0.479} 38.9 & \cellcolor[rgb]{0.108,0.52,0.25} \textcolor{white}{\textbf{62.0}} & \cellcolor[rgb]{0.602,0.837,0.583} 31.7 & \cellcolor[rgb]{0.794,0.919,0.768} 18.8 & 72\\
 & UnboundLocal & \cellcolor[rgb]{0.953,0.982,0.943} 2.3 & \cellcolor[rgb]{0.915,0.968,0.899} 7.6 & \cellcolor[rgb]{0.913,0.967,0.896} 8.0 & \cellcolor[rgb]{0.53,0.804,0.524} \textbf{35.6} & \cellcolor[rgb]{0.931,0.974,0.917} 5.4 & \cellcolor[rgb]{0.951,0.981,0.94} 2.7 & 72\\
 & Value & \cellcolor[rgb]{0.794,0.919,0.768} 18.8 & \cellcolor[rgb]{0.915,0.968,0.899} 7.5 & \cellcolor[rgb]{0.502,0.791,0.501} \textbf{37.4} & \cellcolor[rgb]{0.53,0.804,0.524} 35.9 & \cellcolor[rgb]{0.842,0.938,0.819} 14.8 & \cellcolor[rgb]{0.879,0.953,0.858} 11.8 & 72\\
 & ZeroDivision & \cellcolor[rgb]{0.402,0.742,0.437} 42.8 & \cellcolor[rgb]{0.474,0.778,0.479} 38.8 & \cellcolor[rgb]{0.168,0.578,0.295} \textcolor{white}{57.4} & \cellcolor[rgb]{0.005,0.432,0.176} \textcolor{white}{\textbf{69.6}} & \cellcolor[rgb]{0.491,0.785,0.492} 38.1 & \cellcolor[rgb]{0.208,0.621,0.327} \textcolor{white}{53.9} & 72\\
 & \textsl{No exception} & \cellcolor[rgb]{0.746,0.899,0.72} 22.4 & \cellcolor[rgb]{0.805,0.924,0.78} 17.9 & \cellcolor[rgb]{0.756,0.903,0.729} 21.7 & \cellcolor[rgb]{0.624,0.847,0.602} \textbf{30.5} & \cellcolor[rgb]{0.714,0.886,0.688} 24.6 & \cellcolor[rgb]{0.802,0.922,0.776} 18.4 & 75\\
 & \textsl{Avg (micro)} & \cellcolor[rgb]{0.7,0.88,0.674} 25.4 & \cellcolor[rgb]{0.765,0.907,0.738} 21.0 & \cellcolor[rgb]{0.574,0.824,0.561} 33.4 & \cellcolor[rgb]{0.27,0.678,0.372} \textbf{49.3} & \cellcolor[rgb]{0.704,0.882,0.679} 25.0 & \cellcolor[rgb]{0.794,0.919,0.768} 18.9 & -\\
\midrule
\multirow{8}{*}{Ruby} & Argument & \cellcolor[rgb]{0.842,0.938,0.819} 14.8 & \cellcolor[rgb]{0.802,0.922,0.776} 18.3 & \cellcolor[rgb]{0.924,0.971,0.909} 6.6 & \cellcolor[rgb]{0.765,0.907,0.738} \textbf{21.2} & \cellcolor[rgb]{0.883,0.955,0.862} 11.2 & \cellcolor[rgb]{0.868,0.949,0.846} 12.8 & 114\\
 & FloatDomain & \cellcolor[rgb]{0.904,0.963,0.886} 9.1 & \cellcolor[rgb]{0.969,0.988,0.961} 0.0 & \cellcolor[rgb]{0.82,0.93,0.795} \textbf{16.7} & \cellcolor[rgb]{0.969,0.988,0.961} 0.0 & \cellcolor[rgb]{0.969,0.988,0.961} 0.0 & \cellcolor[rgb]{0.872,0.95,0.85} 12.3 & 8\\
 & Name & \cellcolor[rgb]{0.929,0.973,0.914} 5.9 & \cellcolor[rgb]{0.579,0.827,0.565} 33.0 & \cellcolor[rgb]{0.276,0.681,0.375} 48.8 & \cellcolor[rgb]{0.069,0.487,0.222} \textcolor{white}{\textbf{64.7}} & \cellcolor[rgb]{0.667,0.866,0.643} 27.6 & \cellcolor[rgb]{0.9,0.962,0.881} 9.8 & 114\\
 & NoMethod & \cellcolor[rgb]{0.842,0.938,0.819} 14.9 & \cellcolor[rgb]{0.872,0.95,0.85} 12.2 & \cellcolor[rgb]{0.672,0.868,0.647} 27.4 & \cellcolor[rgb]{0.634,0.852,0.611} \textbf{29.9} & \cellcolor[rgb]{0.816,0.928,0.791} 17.0 & \cellcolor[rgb]{0.794,0.919,0.768} 19.0 & 114\\
 & Range & \cellcolor[rgb]{0.969,0.988,0.961} 0.0 & \cellcolor[rgb]{0.969,0.988,0.961} 0.0 & \cellcolor[rgb]{0.969,0.988,0.961} 0.0 & \cellcolor[rgb]{0.969,0.988,0.961} 0.0 & \cellcolor[rgb]{0.969,0.988,0.961} 0.0 & \cellcolor[rgb]{0.969,0.988,0.961} 0.0 & 3\\
 & Type & \cellcolor[rgb]{0.907,0.964,0.888} 8.8 & \cellcolor[rgb]{0.879,0.953,0.858} 11.6 & \cellcolor[rgb]{0.879,0.953,0.858} 11.7 & \cellcolor[rgb]{0.686,0.874,0.661} \textbf{26.5} & \cellcolor[rgb]{0.839,0.937,0.815} 15.3 & \cellcolor[rgb]{0.868,0.949,0.846} 12.6 & 114\\
 & ZeroDivision & \cellcolor[rgb]{0.552,0.814,0.542} 34.6 & \cellcolor[rgb]{0.351,0.718,0.412} 45.1 & \cellcolor[rgb]{0.168,0.578,0.295} \textcolor{white}{57.5} & \cellcolor[rgb]{0.153,0.562,0.283} \textcolor{white}{\textbf{58.6}} & \cellcolor[rgb]{0.524,0.801,0.52} 36.1 & \cellcolor[rgb]{0.502,0.791,0.501} 37.2 & 52\\
 & \textsl{No exception} & \cellcolor[rgb]{0.868,0.949,0.846} 12.5 & \cellcolor[rgb]{0.775,0.911,0.747} 20.6 & \cellcolor[rgb]{0.791,0.918,0.764} 19.2 & \cellcolor[rgb]{0.704,0.882,0.679} \textbf{25.1} & \cellcolor[rgb]{0.798,0.921,0.772} 18.6 & \cellcolor[rgb]{0.824,0.931,0.799} 16.3 & 51\\
 & \textsl{Avg (micro)} & \cellcolor[rgb]{0.857,0.944,0.835} 13.7 & \cellcolor[rgb]{0.751,0.901,0.724} 21.9 & \cellcolor[rgb]{0.7,0.88,0.674} 25.4 & \cellcolor[rgb]{0.541,0.809,0.533} \textbf{35.0} & \cellcolor[rgb]{0.787,0.916,0.76} 19.4 & \cellcolor[rgb]{0.827,0.933,0.803} 16.2 & -\\

\bottomrule
    \end{tabular}  
    \begin{tablenotes}
	\item[1] Shortened by removing \enquote{Exception} and \textquote{Error} suffixes
    \item[2] CodeGemma
    \item[3] DeepSeek Coder 2
    \item[4] Codestral
    \item[5] Qwen2.5 Coder
    \item[6] DeepSeek Coder
    \item[7] CodeLlama
    \end{tablenotes}

\end{threeparttable}
\end{table}

\section{Evaluation Results}\label{sec:results}
Evaluation results are shown in Tables~\ref{tab:models} and \ref{tab:results}. We will briefly discuss results from three perspectives: model, programming language and exception type.

\paragraph*{Model perspective}
We see in Table~\ref{tab:models} that Qwen2.5 Coder achieves the best performance among the models evaluated. We further see that model size and performance is not strongly correlated (Pearson's $\rho$=0.39). In particular, the 7 billion CodeGemma~\citep{teamCodeGemmaOpenCode2024} fared better than the 34 billion CodeLlama~\citep{roziereCodeLlamaOpen2024}, which, although being the largest model on our list, had the lowest F1 score of all tested models. For CodeGemma we could afford to use a model variant with 16-bit weights (due to its lower parameter count) while all other models are 4-bit quantized. To assess whether this biases results we also evaluated CodeGemma with 4-bit quantization and found no significant performance difference (in fact, the 4-bit version performed 0.2\,\% better).

\paragraph*{Language perspective}
Averaging results over all models, we find that performance is lowest for Ruby (22\,\%) and highest for Python (29\,\%).
Table~\ref{tab:results} shows results broken down by language and exception type (numbers represent F1 scores). We can discern a certain model-specific preference for specific languages. For instance, Qwen2.5 Coder performs best for 9 out 11 Python exceptions, but only for 3 of 10 Java exceptions. When taking the result of the best-performing model for each language, we find very similar results (i.e., an F1 score of roughly 35\,\%), except for Python, where Qwen2.5 Coder obtains an F1 score of almost 50\,\%.

\paragraph*{Exception perspective}
The coloring of Table~\ref{tab:models} reveals that there are certain exception types where all models show decent performance (visible as a green vertical \enquote{band}). 
We see such bands for \texttt{ZeroDevision} exceptions, irrespective of language (note that in Java the corresponding exception is called \texttt{ArithmeticException}). Similarly, white (or light green) bands indicate exceptions that proved difficult for all models. These include, for instance, Ruby's \texttt{RangeError} (which has very low support), overflow errors (e.g., \texttt{OverflowError} in Python or \texttt{StackOverflowException} in Java) as well as null dereference exceptions, such as Java's \texttt{NullPointerException} or \texttt{NullReferenceException} and \texttt{ArgumentNullException} in C\#. Very good performance ($> 0.7$) was obtained on Python exceptions of type \texttt{SyntaxError} and \texttt{AttributeError} as well as \texttt{DivisionByZeroException} in C\#.

\paragraph*{Summary of results}
In summary, we find that \throwbench{} is challenging for current models (the best model obtained an F1 score of 0.38). Moreover, performance is not consistent over languages and exception types. For instance, the best model (Qwen2.5 Coder~\citep{huiQwen25CoderTechnicalReport2024}) performs well on Python, but shows weaknesses on Java, where it is outperformed by DeepSeek Coder~\citep{guoDeepSeekCoderWhenLarge2024}. Division by zero exceptions where recognized with relatively high success; on the other hand, overflow errors or null pointer dereference errors posed a major challenge to all six models.

\section{Conculsions}
In this work we have presented \throwbench{}, a multi-lingual (Python, Ruby, C\# and Java) benchmark for evaluating large code LLMs. In this benchmark, models must predict the type of exception a given program throws upon specific program input (or state that the given program does not throw an exception). Our benchmark is intended as an extension to existing benchmarks in order to obtain a fuller picture of a model's code capabilities. As ground-truth answers (i.e., the exception type thrown) were obtained through program execution we can assume that evaluation results of current models are not biased by possible contamination of training data.

With the best model (Qwen2.5 Coder) obtaining an F1 score of 38\,\%, our evaluation shows that there is ample room for improvements.
The prompt used in our evaluation provide the model with answer options (i.e., a list of possible exception names). In the future, these answer choices could be omitted to make the benchmark even more challenging. Moreover, Chain-of-Though promoting~\citep{weiChainofThoughtPromptingElicits2023} may be a promising technique to obtain better results. We leave this to future work.

\throwbench{} can be downloaded from \downloadurl{}.

\printbibliography{}

\end{document}